# Quantitatively Analyzing Phonon Spectral Contribution of Thermal Conductivity Based on Non-Equilibrium Molecular Dynamics Simulation I: From Space Fourier Transform

Yanguang Zhou[1,¶], Xiaoliang Zhang[2,¶], and Ming Hu[1,2,*]

[1]*Aachen Institute for Advanced Study in Computational Engineering Science (AICES), RWTH Aachen University, 52062 Aachen, Germany*

[2]*Institute of Mineral Engineering, Division of Materials Science and Engineering, Faculty of Georesources and Materials Engineering, RWTH Aachen University, 52064 Aachen, Germany*

## Abstract

Probing detailed spectral dependence of phonon transport properties in bulk materials is critical to improve the function and performance of structures and devices in a diverse spectrum of technologies. Currently, such information can only be provided by the phonon spectral energy density (SED) or equivalently time domain normal mode analysis (TDNMA) methods in the framework of equilibrium molecular dynamics simulation (EMD), but has not been realized in non-equilibrium molecular dynamics simulations (NEMD) so far. In this paper we generate a new scheme directly based on NEMD and lattice dynamics theory, called time domain direct decomposition method (TDDDM), to predict the phonon mode specific thermal conductivity. Two benchmark cases of Lennard-Jones (LJ) Argon and Stillinger-Weber (SW) Si are studied by TDDDM to characterize contributions of individual phonon modes to overall thermal conductivity and the results are compared with that predicted using SED and TDNMA. Excellent agreements are found for both cases, which confirm the validity of our TDDDM approach. The biggest advantage of TDDDM is that it can be used to investigate the size effect of individual phonon modes in NEMD simulations, which cannot be tackled by SED and TDNMA in EMD simulations currently. We found that the phonon modes with mean free path larger than the system size are truncated in NEMD and contribute little to the overall thermal conductivity. The TDDDM provides direct physical origin for the well-known strong size effects in thermal conductivity prediction by NEMD. Moreover, the well-known common sense of the zero thermal conductivity contribution from Gamma point is rigorously proved by TDDDM. Since TDDDM inherently possesses the nature of both NEMD simulation and lattice dynamics, we anticipate that TDDDM is particularly useful for offering deep understanding of phonon behaviors in nanostructures or under strong confinement, especially when the structure size is significantly smaller than the characteristic mean free path of the dominant phonons.

---

[¶] Y.Z. and X.Z. contributed equally to this work.
[*] Author to whom all correspondence should be addressed. E-Mail: hum@ghi.rwth-aachen.de (M.H.)

# I. INTRODUCTION

Understanding the mechanisms of heat transfer, especially at small scales, is of crucial importance to improve the function and performance of devices in many nanotechnology applications [1], such as thermal barrier coatings, novel thermoelectrics (energy conversion), heat sink materials (thermal management) [2], etc. Traditionally, some methods from continuum level have been proposed and applied to predict the phonon mode contribution of thermal conductivity. Klemens [3, 4] and other researchers [5-8] obtained the spectral contribution to thermal conductivity based on long-wave approximation and Debye model. The third-order anharmonic lattice dynamic (ALD) calculation was firstly proposed by Maradudin *et al.* [9] and then extended by Debernardi *et al.* [10] and Deinze *et al.* [11]. Later on, Omini and Sparaviga [12] proposed an iterative scheme to give exact solutions to the linearized BTE and then obtained the lattice thermal conductivity. However, all these methods can only deal with relatively simple systems such as perfect bulk crystalline materials and cannot assess the effect of defects from the real atomic model.

In principle, heat transfer at nanosacle may differ significantly from that in macro- and micro-scales [2, 13]. With characteristic length of devices or structures becoming comparable to or even smaller than the phonon mean free path (MFP) and wavelength of heat carriers (mainly refers to phonons herein), conventional heat transfer theory is no longer valid. Tremendous progress has been made in the past two decades to understand and characterize heat transfer at nanoscales or in nanostructures [14]. From numerical modeling point of view, atomistic simulations such as classical molecular dynamics (MD) and Green's function method, and the phonon-level calculations such as Boltzmann transport equation (BTE) combined with *ab initio*, are quite useful approaches to characterize the intrinsic properties of heat transfer at nanoscale [15-19]. Here, we focus on classical MD method where the phonon-phonon interactions at all levels of orders are described exactly. So far, the two most commonly used approaches to predict the thermal conductivity in MD simulations are the Green-Kubo (GK) method [20,21] and the direct method [22-24]. Usually the former is also called equilibrium molecular dynamics (EMD) simulation and the latter is called non-equilibrium molecular dynamics (NEMD) simulation. Both GK-EMD and NEMD methods can predict the overall thermal conductivity, but give limited insight of the detailed phonon mode contribution. Recently, rigorous calculation of individual phonon mode contribution to the overall thermal transport based on EMD simulation has become available, namely the spectral energy density (SED) [25-30] and equivalently the time domain normal mode analysis (TDNMA) [31-36] methods. Although they have successfully predicted the phonon behavior in perfect bulk systems, in particular mode specific phonon lifetime, both methods need some assumptions to obtain the modal thermal conductivity when they are deal with complicated systems, such as materials with interfaces [37] or disorder materials [36]. Furthermore, fitting the Lorentzian function in SED and the normalized autocorrelation function of the mode energy in TDNMA is sometimes nontrivial and may bring significant errors to the final result.

In this paper, we propose a new method named time domain direct decomposition method (TDDDM), which is directly based on NEMD simulations, to calculate the contributions of individual phonon mode contribution to the overall thermal conductivity. Thus, in principle the TDDDM can be used as a tool to study the phonon behavior in nanostructures or the length of system along which the phonon travel is limited. It was well documented in literature that the size effect is very important when the average mean free path (MFP) is comparable to the system size [22]. Using the new method, we are able to investigate the size effect of individual phonon mode contribution, which cannot be realized by the previous SED and TDNMA methods that are based on EMD simulation. We find that the phonon modes with mean free path larger than the system size are truncated. Meanwhile, the thermal conductivity contributed by the Gamma point is rigorously proved to be zero by our TDDDM approach directly.

The remainder of this paper is organized as follows. In Sec. II, we present the theoretical framework of our TDDDM method and the previous methods (TDNMA and SED) to characterize the individual phonon mode contribution to the overall thermal conductivity. The implementation details of the TDDDM scheme are given in Sec. III. We then present the validation of the TDDDM framework from the aspect of heat current of phonon modes as well as phonon mode specific thermal conductivity contribution in Sec. IV, followed by the discussions on some important computational parameters in the TDDDM framework in Sec. V. In Sec. VI, we apply the TDDDM to study phonon mode contributions in finite size systems and the thermal conductivity contribution from the Gamma point. In Sec. VII, we present a summary and conclusion.

## II. CHARACTERIZING INDIVIDUAL PHONON MODE CONTRIBUTION TO OVERALL THERMAL CONDUCTIVITY

### A. Framework based on non-equilibrium molecular dynamics simulation

The molecular dynamics expression of heat current was first derived by Irving and Kirkwood [38], and then extended by Daichi et al. [39, 40] The detailed formula of heat current can be written as

$$\mathbf{Q} = \left\langle \sum_j (E_j \mathbf{v}_j - \mathbf{S}_j \mathbf{v}_j) \right\rangle, \quad (1)$$

where $\mathbf{Q}$ is the heat current, the sum is taken over all atoms in the interested volume, $\langle \rangle$ denotes the time average, $E_j$ and $\mathbf{S}_j$ are the total energy and symmetric second order stress tensor of atom $j$, respectively. The detailed expression of $\mathbf{S}_j$ can be found in Ref. [41].

According to the theory of lattice dynamics, we can write the displacement equation of any atom as [42]

$$\mathbf{u}(jl,t) = \sum_{\mathbf{k},\nu} \mathbf{U}(j,\mathbf{k},\nu) \exp(i[\mathbf{k}\mathbf{r}(jl,) - \omega(\mathbf{k},\nu)t]), \quad (2)$$

where the sum is taken over all the wave vectors $\mathbf{k}$ and all phonon branches $\nu$, $\mathbf{U}(j,\mathbf{k},\nu)$ is the amplitude vector that tells us how the atom $j$ in the unit cell $l$ moves induced by the phonon mode $(\mathbf{k}, \nu)$, which gives the direction and the amplitude of the motion, and $\omega(\mathbf{k}, \nu)$ is the vibrational frequency of the phonon mode $(\mathbf{k}, \nu)$.

By applying space inverse Fourier transform to Eq. (2), we can obtain the normal mode coordinate of each vibrational mode

$$X(\mathbf{k},\nu;t) = \frac{1}{N^{1/2}} \sum_{jl} m_j^{1/2} \exp(-i\mathbf{k}\mathbf{r}(jl))\mathbf{e}^*(j,\mathbf{k},\nu)\mathbf{u}(jl,t), \qquad (3)$$

where $N$ is the total number of unit cells and $\mathbf{e}$ is the mode eigenvector, which can be obtained by diagonalizing the dynamic matrix $\mathbf{D(k)}$.

After that, based on Fourier transform, the displacement of each atom can be written using the normal mode coordinate $X(\mathbf{k},\nu)$ as

$$\mathbf{u}(jl,t) = \frac{1}{(Nm_j)^{1/2}} \sum_{\mathbf{k},\nu} \mathbf{e}(j,\mathbf{k},\nu) \exp(i\mathbf{k}\mathbf{r}(jl)) X(\mathbf{k},\nu;t). \qquad (4)$$

Since $E_j$ and $\mathbf{S}_j$ in Eq. (1) are related to the displacement of atom $j$ [17, 43, 44], their space Fourier transform, denoted by $\widetilde{E}$ and $\widetilde{\mathbf{S}}$, respectively, can be defined as

$$\widetilde{E}(\mathbf{k},\nu;t) = \frac{1}{N^{1/2}} \sum_{jl} \xi_1(j,\mathbf{k},\nu) \exp(i\mathbf{k}\mathbf{r}(jl)) E(jl,t), \qquad (5a)$$

$$\widetilde{\mathbf{S}}(\mathbf{k},\nu;t) = \frac{1}{N^{1/2}} \sum_{jl} \xi_2(j,\mathbf{k},\nu) \exp(i\mathbf{k}\mathbf{r}(jl)) \mathbf{S}(jl,t), \qquad (5b)$$

where the $\xi_1$ and $\xi_2$ represents the polarization vector of $E$ and $\mathbf{S}$, respectively.

Then, $E_j$ and $\mathbf{S}_j$ can be rewritten as

$$E(jl,t) = \frac{1}{N^{1/2}} \sum_{jl} \xi_1^*(j,\mathbf{k},\nu) \exp(-i\mathbf{k}\mathbf{r}(jl)) \widetilde{E}(\mathbf{k},\nu;t), \qquad (6a)$$

$$\mathbf{S}(jl,t) = \frac{1}{N^{1/2}} \sum_{jl} \xi_2^*(j,\mathbf{k},\nu) \exp(-i\mathbf{k}\mathbf{r}(jl)) \widetilde{\mathbf{S}}(\mathbf{k},\nu;t). \qquad (6b)$$

Using Eqs. (1), (4), (6a) and (6b), we can write the heat current contributed by each atom, $\mathbf{q}(jl)$, in the wave-vector reciprocal space as

$$\begin{aligned}
\mathbf{q}(jl) &= \langle (E(jl) - \mathbf{S}(jl)) \mathbf{v}(jl) \rangle \\
&= \left\langle \frac{1}{N^{1/2}} \left\{ \sum_{\mathbf{k},\nu} \xi_1^*(j,\mathbf{k},\nu) \exp(-i\mathbf{k}\mathbf{r}(jl)) \widetilde{E}(\mathbf{k},\nu) - \right. \right. \\
&\quad \left. \sum_{\mathbf{k},\nu} \xi_2^*(j,\mathbf{k},\nu) \exp(-i\mathbf{k}\mathbf{r}(jl)) \widetilde{\mathbf{S}}(\mathbf{k},\nu) \right\} \cdot \\
&\quad \left. \frac{1}{(Nm_j)^{1/2}} \sum_{\mathbf{k},\nu} \mathbf{e}(j,\mathbf{k},\nu) \exp(i\mathbf{k}\mathbf{r}(jl)) \dot{X}(\mathbf{k},\nu) \right\rangle
\end{aligned} \qquad (7)$$

Here, we define the contribution of an individual phonon mode $(\mathbf{k},\nu)$ to the heat current of an individual atom ($jl$) as

$$\begin{aligned}
\mathbf{q}(jl;\mathbf{k},\nu) &= \left\langle \frac{1}{N^{1/2}} \left\{ \sum_{\mathbf{k},\nu} \xi_1^*(j,\mathbf{k},\nu) \exp(-i\mathbf{k}\mathbf{r}(jl)) \widetilde{E}(\mathbf{k},\nu) - \right. \right. \\
&\quad \left. \sum_{\mathbf{k},\nu} \xi_2^*(j,\mathbf{k},\nu) \exp(-i\mathbf{k}\mathbf{r}(jl)) \widetilde{\mathbf{S}}(\mathbf{k},\nu) \right\} \cdot \\
&\quad \left. \frac{1}{(Nm_j)^{1/2}} \mathbf{e}(j,\mathbf{k},\nu) \exp(i\mathbf{k}\mathbf{r}(jl)) \dot{X}(\mathbf{k},\nu) \right\rangle
\end{aligned} \qquad (8)$$

Thus, the heat current of each phonon mode $\mathbf{Q}(\mathbf{k},\nu)$ can be described as

$$\begin{aligned}
\mathbf{Q}(\mathbf{k},\nu) &= \sum_{jl} \left\langle \frac{1}{N^{1/2}} \left\{ \sum_{\mathbf{k},\nu} \xi_1^*(j,\mathbf{k},\nu) \exp(-i\mathbf{k}\mathbf{r}(jl)) \widetilde{E}(\mathbf{k},\nu) - \right. \right. \\
&\quad \left. \sum_{\mathbf{k},\nu} \xi_2^*(j,\mathbf{k},\nu) \exp(-i\mathbf{k}\mathbf{r}(jl)) \widetilde{\mathbf{S}}(\mathbf{k},\nu) \right\} \cdot \\
&\quad \left. \frac{1}{(Nm_j)^{1/2}} \mathbf{e}(j,\mathbf{k},\nu) \exp(i\mathbf{k}\mathbf{r}(jl)) \dot{X}(\mathbf{k},\nu) \right\rangle \\
&= \sum_{jl} \left\langle (E(jl,t) - \mathbf{S}(jl,t)) \frac{1}{(Nm_j)^{1/2}} \mathbf{e}(j,\mathbf{k},\nu) \exp(i\mathbf{k}\mathbf{r}(jl)) \dot{X}(\mathbf{k},\nu) \right\rangle
\end{aligned} \qquad (9)$$

Finally, by assuming the same temperature gradient for all phonon modes, the contribution of an individual phonon mode $(\mathbf{k},\nu)$ to the overall thermal conductivity can be calculated by Fourier's law

$$\mathbf{K}_{\mathrm{TDDDM}}(\mathbf{k},\nu) = -\frac{1}{V_c} \frac{\mathbf{Q}(\mathbf{k},\nu)}{\nabla T}, \qquad (10)$$

where $V_c$ is the volume of control box in a NEMD simulation (see Fig. 1 in Sec. III for details). Here, the direct results from TDDDM are the decomposed heat current. We should combine Eq. (10) and Eq. (11) if we want to know more information about phonons, such as relaxation time and MFP.

### B. Framework based on equilibrium molecular dynamics simulation

Using the relaxation time approximation[45] to solve Boltzmann transport equation yields an expression of thermal conductivity $\mathbf{K}_{\mathrm{BTE}}(\mathbf{k},\nu)$ [46]

$$\mathbf{K}_{\text{BTE}}(\mathbf{k},\nu) = c_{ph}(\mathbf{k},\nu)\mathbf{v}_g^2(\mathbf{k},\nu)\tau(\mathbf{k},\nu),\qquad(11)$$

where the phonon mode $(\mathbf{k},\nu)$ has group velocity $\mathbf{v}_g(\mathbf{k},\nu)$, volume specific heat $c_{ph}(\mathbf{k},\nu)$, and phonon lifetime $\tau(\mathbf{k},\nu)$. In classical molecular dynamics simulation, the volumetric specific heat can been obtained by $c_{ph}(\mathbf{k},\nu) = k_b/V$, where $k_b$ and V is the Boltzmann's constant and system volume, respectively.

In the framework of time domain, the phonon mode lifetime $\tau_{\text{TDNMA}}(\mathbf{k},\nu)$ can be calculated using TDNMA (run in EMD) [31]

$$\tau_{\text{TDNMA}}(\mathbf{k},\nu) = \int_0^{t^*}\left(\frac{\langle E(\mathbf{k},\nu;t)E(\mathbf{k},\nu;0)\rangle}{\langle E(\mathbf{k},\nu;0)E(\mathbf{k},\nu;0)\rangle}\right)dt,\qquad(12)$$

where the upper integration limit $t^*$ should be much longer than the lifetime of the specific phonon. The total energy of each phonon mode, $E(\mathbf{k},\nu)$, is calculated from

$$E(\mathbf{k},\nu;t) = \frac{1}{2}\omega^2(\mathbf{k},\nu)X(\mathbf{k},\nu;t)X^*(\mathbf{k},\nu;t) + \frac{1}{2}\dot{X}(\mathbf{k},\nu;t)\dot{X}^*(\mathbf{k},\nu;t)\qquad(13)$$

In contrast, in the framework of frequency domain (also run in EMD), a spectral energy density (SED), $\Phi(\mathbf{k},\nu)$, can be calculated by [26]

$$\Phi(\mathbf{k},\nu) = 2\lim_{t_0\to\infty}\frac{1}{2t_0}\left|\frac{1}{\sqrt{2\pi}}\int_0^{t_0}X(\mathbf{k},\nu;t)\exp(-i\omega t)dt\right|^2.\qquad(14)$$

Fitting the SED using the Lorentzian function with center at $\omega_0(\mathbf{k},\nu)$ and line width equal to $\Gamma(\mathbf{k},\nu)$, one can obtain

$$\Phi(\mathbf{k},\nu) = C_0(\mathbf{k},\nu)\frac{\Gamma(\mathbf{k},\nu)/\pi}{[\omega_0(\mathbf{k},\nu)-\omega]^2 + \Gamma^2(\mathbf{k},\nu)}, \quad (15)$$

where $C_0(\mathbf{k},\nu)$ is a mode dependent constant. The lifetime of the phonon mode $\tau_{SED}(\mathbf{k},\nu)$ is then given by [31]

$$\tau_{SED}(\mathbf{k},\nu) = \frac{1}{2\Gamma(\mathbf{k},\nu)}. \quad (16)$$

Finally, the thermal conductivity contributed by each phonon mode can be calculated from Eq. (11). Here, it is worth noting that the thermal conductivity computed by Eq. (11) is a tensor. In our calculation we only calculated the thermal conductivity along the concerned direction.

### III. IMPLEMENTATION OF TIME DOMAIN DIRECT DECOMPOSITION METHOD

#### A. Set up TDDDM simulation system

The TDDDM described in Sec. II A is based on direct-MD method (NEMD), where the thermal conductivity is evaluated by imposing a constant heat flux across the model and measuring the resulting temperature gradient, or alternatively, by applying a constant temperature gradient and calculating the resulting heat flux. A schematic of a NEMD simulation is illustrated in Fig. 1. Fixed boundary conditions are applied to both ends of the system. Near to the fixed boundaries, a hot and cold reservoir is applied such that a temperature gradient is established along the concerned direction ($z$ direction in our simulations) after running some time. Periodic boundary conditions are applied in the lateral ($x$ and $y$) directions. For all of our LJ Argon systems, the fixed boundaries contain 4-unit-cell long Ar atoms, and the hot and cold reservoirs include 12-unit-cell long Ar atoms. While for the models of SW Si, 4-unit-cell long and 16-unit-cell long Si atoms are used as the fixed boundaries and reservoirs, respectively. Regarding the interatomic potential, we used the standard 12-6 Lennard-Jones (LJ) potential [47] for solid argon at 10 K and the original Stillinger-Weber (SW) potential [48] for Si at 300 K. It is well known that there exist size effects for thermal conductivity in NEMD simulations [22, 49], which can be investigated by varying the length of the system along the concerned direction. Here, we choose our simulation systems as $8a_{Ar} \times 8a_{Ar} \times n \cdot a_{Ar}$ for LJ Argon and $4a_{Si} \times 4a_{Si} \times n \cdot a_{Si}$ for SW Si ($a$ is the lattice parameters, i.e. 0.529 nm and 0.544 nm for LJ Argon and SW Si, respectively), where $n$ ranges from 200 to 500 for LJ Argon and from 500 to 4000 for SW Si. So, the total length of our simulation systems in $z$ direction is about 100 – 270 nm for LJ Argon and 250 – 2000 nm for SW Si. There could be size effect in the $y$ and $z$ directions since the model only contains 4 or 8 unit cells in the cross section direction. However, we have proved that the size effect of the cross section on both the modal and total thermal conductivity can be ignored, by applying the periodic boundary conditions along the lateral direction. Each simulation run starts with 2 ns *NPT* (constant particles, pressure, and temperature) and *NVE* (constant particles, volume and no thermostat) relaxation to allow the system to reach equilibrium state.

Following equilibration, we computed the thermal conductivity of the system using the NEMD method. The temperature gradient of the model systems is kept to be a constant (about 0.008 K/$a_{Ar}$ and 0.0075 K/$a_{Si}$ for LJ Argon and SW Si, respectively) by using the Langevin thermostat. Therefore, at each time step a heat ($\Delta\varepsilon$) is added into the hot reservoir and the same amount of heat is removed from the cold reservoir. In our MD simulations, the coupling strength parameter ($t_{bath}$) of the Langevin thermostat is chosen as 0.1 ps. It has been shown in previous studies that the calculated thermal conductivity is independent of the length of the reservoirs, when the bath-induced mean free path of phonons satisfies $\Lambda_{bath} = c_s \cdot t_{bath} \ll L_{bath}$[46, 50], where $c_s$ is the speed of sound (~ 1250 m/s for Argon and 9620 m/s for Si). The first ~ 10 ns NEMD run for LJ Argon (10 - 20 ns for SW Si, depending on system length) is used to obtain the steady heat current, then the mode specific heat current contribution is extracted from the later on 6 ns run.

### B. Choose appropriate size of control volume

From the framework of TDDDM we know that TDDDM is not only based on NEMD simulation, but also based on the lattice dynamics (LD) theory. Therefore, the fundamental of LD should be valid in the implemented volume in our simulations. Here, the volume to implement the TDDDM approach is named as "control volume", as shown in Fig. 1. There are three main rules to choose the size of the control volume to make sure that the LD is valid in our NEMD simulations. Firstly, the temperature difference along the sample should be very small (< 0.1 K in all our simulations) due to the steady heat flux. This value is actually much less than the intrinsic temperature fluctuation. Secondly, the distance between the control volume and the thermostat reservoir should be far enough, such that the nonlinear effect of the reservoir can be neglected. In other words, the position of the control volume should be in the linear region of the temperature gradient. Lastly, the computational efficiency should be considered as well, since the computational cost of TDDDM is a bit expensive. In our simulations, the size of the control volume of LJ Argon and SW Si is $8a_{Ar} \times 8a_{Ar} \times 8a_{Ar}$ and $4a_{Si} \times 4a_{Si} \times (4 - 16)a_{Si}$, respectively. Thus, the temperature difference along the control volume for Argon and Si is about 0.06 and 0.02 – 0.08 K, respectively. Here, the temperature difference is calculated by fitting the average temperature profile. The shortest distance between the control volume and the thermostat reservoir is $92a_{Ar}$ (LJ Argon) and $240a_{Si}$ (SW Si), which are far enough to avoid the non-linear effect of the reservoir. Moreover, based on lattice dynamics theory, the number of extractable phonon modes is determined by the total number of atoms in the control volume ($3N$ phonon modes exist for the control volume with $N$ atoms). Thus, the number of phonon modes can be extracted in our system are 6144 for LJ Argon and 1536 – 6144 for SW Si. One point we also need to mention is that we use a reduced number of modes due to the limited control volume to implement the TDDDM, and thus the absolute values of phonon properties (such as relaxation time and MFP) have some deviation from their real values. Enlarging the control volume or in other words increasing the number of extractable phonon modes can reduce such effect.

## C. Computational details of SED/TDNMA simulations

In order to validate our TDDDM results, we also run EMD simulations to extract phonon properties using SED and TDNMA techniques. The supercells of the simulation systems in SED/TDNMA are chosen to be equal to the control volume in TDDDM. Here, the size of models in SED/TDNMA is chosen as $8a_{Ar} \times 8a_{Ar} \times 8a_{Ar}$ (LJ Argon at 10 K) and $4a_{Si} \times 4a_{Si} \times 4a_{Si}$ (SW Si at 300 K) with periodic boundary conditions in all directions. Since our model systems do not contain defects and internal interfaces, it is not surprising to find that, the largest MFP reaches 65 nm and 1420 nm for LJ Argon and SW Si, respectively. In real samples, which involve additional phonon scattering mechanisms, such as impurities, the largest MFP is usually much shorter [51]. According to the previous discussion, the size of TDDDM models in $z$ direction should be larger than 1420 nm, such that all the phonon modes mentioned above can appear in TDDDM. Thus, the size of the corresponding models in TDDDM is chose as $8a_{Ar} \times 8 a_{Ar} \times ( > 123a_{Ar} )$ for LJ Argon and $4a_{Si} \times 4a_{Si} \times (> 2610a_{Si})$ for SW Si. For SED/TDNMA all the simulations start with 2 ns *NPT* and *NVE* relaxation to allow the system to reach equilibrium state. Setting up the appropriate total running time for SED is related to the cutoff frequency of the two materials. For LJ Argon the cutoff frequency is about 2 THz and for SW Si it is about 18 THz. To include the contributions from all the phonon modes, we need to determine the time interval for SED velocity output according to $f_{max} = 0.5/dt$, where $f_{max}$ is the maximum frequency that SED can reach and $dt$ is the time interval for output. Based on the above formula and the cutoff frequency of each material, we can choose the output interval of Argon and Si to be 0.2 ps and 0.025 ps, respectively. In order to obtain accurate results, for both Argon and Si we use 100, 000 sets of velocity data to perform the SED calculations. Then the total running time for Argon and Si should be at least 20 ns and 2.5 ns, respectively. In contrast, we run about 7 ns for LJ Argon and SW Si to obtain the relaxation time of all phonon modes using TDNMA. Here, one thing we have to mention is that the running time of our SED/TDNMA simulations may be much larger than the convergence time.

The computational cost of each prediction method is also a major concern. The TDNMA method scales as $(Nn)^2$, where $N$ and $n$ represents the number of unit cell and basis atoms in the unit cell, respectively. The computational costs of TDDDM and SED method are proportional to $2(Nn)^2$. Furthermore, for SED and TDNMA which are based on EMD with periodic boundary conditions, the relative small systems can be used. For TDDDM which is based on NEMD, a quite large size along the concerned direction should be adopted, such that all the extractable phonon modes from the corresponding SED/TDNMA simulations can be included. It is also worth noting that it requires a much longer time (about 20 ns) to ensure all the heat current of phonon modes converge. Due to computational cost, the number of phonon modes which can be tackled by TDDDM is limited. Another important issue is the storage space. Since all the three methods mentioned above are post-processing methods. A quite large space (over 10 GB) is required to save the raw data. Here, we embed the TDNMA and TDDDM methods into the open source software LAMMPS and therefore the required space to save the raw data is decreased to hundreds of MBs.

In addition, it is worth pointing out that, all the three methods used in our paper are implemented in finite temperature systems. The effect of the system temperature on the vibration of phonons is considered by using the quasi-harmonic lattice dynamics and the detailed information can be found in Ref. [ 46]. All the results in this paper are averaged over five independent runs, each with different initial velocities by assigning different random seed.

**IV. VALIDATION OF TIME DOMAIN DIRECT DECOMPOSITION METHOD**

**A.   Heat current fluctuation of individual phonon mode**

Since the heat current of each phonon mode in principle is the decomposition of the total heat current, it should also reach the steady state finally as total heat current does. As formulized in Sec. II A, we start the validation of our TDDDM approach with the investigation of the time development of heat current contributed by individual phonon modes. Here, the heat current contributions of three typical individual phonon modes for LJ Argon are used as examples for analysis (Fig. 2). It can be clearly seen from Fig. 2 that, the heat current of the individual phonon mode converges after averaging over about 3 ns. In addition, it is interesting to find that the longitudinal phonon mode (red line) is more stable than the other two transverse phonon modes (green and black line). Using the method mentioned in Sec. III C, we can find that the relaxation time of phonon modes corresponding to red line, green line and black line are 6.27, 20.78 and 17.38 ps, respectively. The longitudinal phonon mode (red line) has smaller relaxation time, which means during the statistic window it appears and disappears more frequently than the other two phonon modes (green and black line). Therefore, the heat current corresponding to this longitudinal phonon mode (red line) is averaged by more times during the same time period, which makes the fluctuation smaller. Furthermore, it is found that the total heat current of the system reaches steady state more quickly. This is understamble considering that the total heat current of the system is the sum of the heat current from all the phonons. The fluctuations of each individual phonon modes may be cancelled out with each other, so the fluctuation of total heat current is much smaller than those of a single phonon mode.

For SW Si, the contribution to the heat current from each phonon mode possesses similar properties. We did not show the results for brevity. Using TDNMA/SED, we can know that the relaxation time of SW Si is larger than that of LJ Argon. It means that the total run time, that is needed to make the heat current of those low frequency phonon modes to be stable in TDDDM, should be quite long. For some cases a total run time of 28 ns is needed for SW Si in our TDDDM simulation. Generally speaking, the lower intrinsic relaxation time a phonon mode has, the shorter run time the corresponding heat current to reach a steady state takes.

In addition, since our TDDDM scheme is established on decomposing the atomic velocity in the heat current formula onto each phonon mode, our heat current computed using TDDDM should in principal be equal to the value calculated by Eq. (1). We have proved this using MD simulations.

### B. Contribution of individual phonon modes to overall thermal conductivity

In this section, the thermal conductivity contribution from each individual phonon modes is predicted by TDDDM and the results are compared with that obtained by SED and TDNMA. Based on Sec. III C, the size of the TDDDM models discussed in this section is chosen as $8a_{Ar} \times 8a_{Ar} \times 500a_{Ar}$ for LJ Argon and $4a_{Si} \times 4a_{Si} \times 4000a_{Si}$ for SW Si. Isotropic approximation is used by Henry and Chen [29] and McGaughey and Kaviany [33] for SW Si and LJ Argon to compute the thermal conductivity contribution of individual phonon modes. However, Turney et. al [46] and Larkin et. al [36] found that the predicted thermal conductivity by an isotropic approximation is lower than that predicted using the full Brillouin zone. Therefore, the full Brillouin zone should be used to calculate the thermal conductivity contributions. To improve the computational efficiency, in this paper the crystal lattice's irreducible Brillouin zone is used to compute the contribution of each phonon mode, which has been shown to predict the same thermal conductivity of the material with respect to use the full Brillouin zone [36]. The results are reported in Fig. 3 (LJ Argon) and Fig. 4 (SW Si). Both figures show that the TDDDM approach predicts identical thermal conductivity contribution of each phonon mode and the accumulative thermal conductivity with respect to mode frequency as well, where the accumulative thermal conductivity is calculated by $K(\omega_0) = \sum_{\omega < \omega_0} K(\omega)$. These results prove that our TDDDM approach yields the same correct results as SED and TDNMA and captures the phonon properties accurately in NEMD runs.

From the left panel of Fig. 3 it is also interesting to see that some phonon modes in the low frequency range of LJ Argon (< 0.5 THz) contribute much more than those in higher frequency region. This phenomenon may lead to the common assumption that low frequency phonon modes dominate the thermal transport [2, 13]. However, from the right panel of Fig. 3, we know that the phonon modes in the medium frequency range (0.7 – 1.5 THz) contribute as much as 70% of the total thermal conductivity. Thus, this common assumption should be considered cautiously (at least invalid for LJ Argon).

For SW Si at 300 K, the acoustic phonon modes contribute over 90% of the predicted thermal conductivity (Fig. 4). It is often assumed that the optical modes ($\omega > 12$ THz for SW Si) do not contribute thermal conductivity considerably, due to their short relaxation time. Here, in our simulations, we find that the optical phonon modes contribute only ~6% to the predicted thermal conductivity, which is in accordance with the previous results reported by Sellan *et al.* [51] and Goicochea *et al.* [34]. Thus, it is reasonable to neglect their contribution to the overall thermal conductivity. Once again, we find that in SW Si some low frequency phonon modes contribute extremely large portion to the overall thermal conductivity as compared to other higher frequency phonon modes. In addition, the accumulative thermal conductivity of the predicted thermal conductivity in the low frequency region (0 - 8 THz) reaches 90%. Therefore, we can conclude that the low frequency phonon modes dominate the thermal transport in bulk Si at room temperature.

What is also worth noting is the absolute value of thermal conductivity. For LJ Argon the absolute thermal conductivity calculated by TDDDM and SED/TDNMA is 3.29 and 2.55 W/mK, respectively, while GK-MD shows the result of such system is 4.00 W/mK [32] and about 2.50 W/mK [36]. For SW Si the thermal conductivity calculated by TDDDM and SED/TDNMA is 381.24 W/mK and 355.32 W/mK (379.66 W/mK in Ref. [34]), respectively. It is obvious that the absolute values of thermal conductivity calculated by TDDDM and TDNMA/SED are different. The reason for this discrepancy could come from the inherent difference between EMD and NEMD and the finite extractable phonon modes in TDNMA/SED.

## V. DISCUSSIONS

In this section, we discuss some important computational parameters that affect the quality of the simulation results by using our TDDDM approach.

### A. Run time of TDDDM simulations

As discussed in Sec. IV A, the total run time in NEMD simulations strongly depends on how fast the heat current of individual phonon modes reaches steady state. It is not surprising to find that, sometimes over a certain period the averaged heat current of some individual phonon modes can even be negative. We are sure that the heat current of all the phonon modes will be positive definitely and will be stabilized after enough long run, since all the phonon modes appearing in the system should contribute to the thermal conductivity, i.e. correspond to a positive heat current. For LJ Argon, it is easy to make all the phonon modes reach a steady state quickly, since the relaxation time of the majority of phonon modes is on the order of $O(1) - O(10)$ ps. In all our simulations of LJ Argon, the largest relaxation time of the extractable phonon modes is about 80.87 ps. While for SW Si, the largest relaxation time of the detected phonon modes in our simulations can reach about 600 ps. As mentioned in Sec. IV A, it takes a quite long time ($> 30$ ns) for such phonon mode to reach steady state. In other words, only over such long period can we make statistics of such phonon modes for several times. Considering the computational efficiency, here, we assume that the heat currents of all the phonon modes are stable, when over 96% of the entire phonon modes reach the steady state. Thus, the total heat current will decrease a little bit (within 3%) with respect to the real heat current. The small difference between the result of TDDDM and that of SED/TDNMA (Fig. 4) may be caused by neglecting the phonon modes with extremely long lifetime in TDDDM.

### B. Effect of the size of control volume

From Sec. III B, we know that the size of the control volume in TDDDM is related to the total number of extractable phonon modes in the system. Here, the SW Si model with system size of $4a_{Si} \times 4a_{Si} \times 2000a_{Si}$ is taken as an example and the size of the control volume varies from $4a_{Si}$ to $16a_{Si}$. As mentioned above, the number of atoms in the control volume increases when the size of the control volume is enlarged, and therefore, more phonon modes can be extracted when the size of the control volume is increased (Fig. 5). As assumed in Sec. II A, we regard that all the phonon modes which can be extracted based on LD contribute to the thermal conductivity. Here, as shown in Fig. 5, the total number of phonon modes can be extracted is 1536 (blue), 4608 (dark purple) and 6144 (light purple), respectively, i.e. larger control volume size corresponds to more datapoints. From Fig. 5 we can see that increasing the size of the control volume only makes the MFP dependent accumulative thermal conductivity smoother due to more datapoints in the curve, but the trend does not change at all. That is to say, the size of the control volume does not affect the longest MFP phonon that can be extracted from TDDDM run, but only determines the quality (smoothness) of the MFP dependent accumulative thermal conductivity.

Using SED/TDNMA, we obtain the range of MFP corresponding to the three control volume sizes to be 0 – 1420 nm, 0 – 1650 nm and 0 – 2130 nm, respectively. It is interesting to observe that for our TDDDM approach only the phonon modes with MFP below ~ 1088 nm (we call it characteristic size) contribute to the thermal conductivity. Phonon modes with MFP larger than 1088 nm are truncated, and therefore, contribute little to the total thermal conductivity. The reason will be explained in detail in Sec. VI A. It is also worth noting that the characteristic size is more accurate and easier to be found when the control volume is larger, since more phonon modes can be extracted in TDDDM run. Another interesting result is that the characteristic size identified in TDDDM is more or less equal to the total size of the simulation system, i.e. $2000a_{Si}$, (Figs. 1 and 5). Thus, we think the phonon modes with MFP larger than the characteristic size (system size) are truncated, which is the physical origin of the size effect in the NEMD simulation (see details in Sec. VI A).

### C. Effect of the system size in TDDDM

As mentioned earlier, TDDDM is directly based on NEMD simulation. Therefore, there is inevitable size effect. Based on the results in last section, the phonon modes with MFP larger than system size will be truncated. For LJ Argon, since the largest phonon MFP is only 65 nm from our SED/TDNMA simulations ($8a_{Ar} \times 8a_{Ar} \times 8a_{Ar}$), It is easy to increase the system size to reach the order of $O(10^2)$ nm. In our TDDDM run, the size of all models along the concerned direction is larger than the largest MFP. Thus, it is straightforward to find that almost all the phonon modes that can be extracted in our TDDDM run contribute to the thermal conductivity. While for SW Si, the largest MFP can be as large as several micrometers [51], which is unreachable for NEMD simulations. When SW Si is simulated using TDDDM, it is easy to observe that many phonon modes are truncated by the finite size of the system (Figs. 5 and 6). The details of the size effect in TDDDM are given in the next section.

## VI. APPLICATIONS

### A. Size effect of thermal conductivity contribution of individual phonon modes

It is well known that, for many materials there exists size effect in thermal conductivity predictions when the NEMD method is used. Since TDDDM is based on the NEMD method, there is also size effect in the phonon mode contribution to the overall thermal conductivity. For LJ Argon, the largest MFP in our model system extracted by SED/TDNMA is about 65 nm. It means that all the intact phonon modes predicted by SED/TDNMA can only appear when the length in $z$ direction is larger than 65 nm. When the system size is smaller than 65 nm, the phonons with MFP larger than 65 nm will be truncated and then become incomplete phonons. It should be noted that the truncation of the long MFP phonons is different from the non-existing of phonons. Here, the dimension in the concerned direction in our models for LJ Argon is chosen to be about 34 – 158 nm. The results of TDDDM for different total length of the system are compared in the left panel of Fig. 6. We find that all the extractable phonon modes contribute to thermal conductivity for the systems with total size in $z$ direction larger than 65 nm. When the system size in $z$ direction is smaller than 65 nm, only phonon modes with MFP smaller than the system size can contribute to the overall thermal conductivity. For SW Si, previous studies have demonstrated that the thermal conductivity computed using NEMD can be seriously affected by the size of the model [21, 22, 49, 51]. The difference in thermal conductivity between NEMD and EMD method can be as large as a few times [21, 49, 52, 53]. In our SED/TDNMA simulations, the largest phonon MFP of SW Si reaches 1420 nm, which is already a quite large size for NEMD simulations. In our TDDDM (NEMD) simulation models, the system size in *z* direction ranges from 272 to 2176 nm. Therefore, the longest simulation size should accommodate the largest MFP phonons. The results are reported in the right panel of Fig. 6. It clearly shows that all the phonon modes contribute to the thermal conductivity when the length in *z* direction is 2176 nm. For the model with length of 1088 nm in *z* direction, phonon modes with MFP larger than about 1 μm are truncated and therefore contribute little to the overall thermal conductivity. The corresponding critical MFP for models with 544 nm in *z* direction is found to be about 500 nm. When the size in *z* direction is decreased to 272 nm (by one order of magnitude smaller than the longest length of 2176 nm), it is interesting to find that the phonon modes with MFP larger than 260 nm (more or less equal to the system size of 272 nm) contribute very little to the overall thermal conductivity, i.e. the corresponding heat current contribution is quite small for these phonon modes. With total length increasing, phonon modes with MFP longer than the system length are present and then they are able to contribute to the heat conduction. Our TDDDM approach provides direct evidence that, the phenomenon that the thermal conductivity of NEMD simulations increases with length increasing is solely resulted from the truncation of long MFP phonons induced by the finite length of the system. Although general understanding to the size effect in NEMD simulation has been established for a long time, it is the first time that physically meaningful and direct evidence is presented and the length dependence of thermal conductivity in NEMD simulation is *quantified* by our TDDDM approach. Certainly, it should be noted that, when TDDDM is implemented to analyze systems with thermal conductivity dominated by very long MFP phonons, both system length and run time should be very long.

**B. Thermal conductivity contribution of Gamma point**

Another interested point in phonon mode specific thermal conductivity is the contribution of Gamma point. The Gamma point corresponds to the bulk translational movement and therefore should not contribute to the thermal conductivity [16]. However, neither SED nor TDNMA can prove this directly. Instead, both methods just *assume* that the Gamma point's contribution is zero. In TDDDM, the heat current of phonon modes corresponding to the Gamma point, i.e. $\mathbf{k} = (0, 0, 0)$, can be explicitly written as [Eq. (9)]

$$\mathbf{Q}(\mathbf{k}=\mathbf{0},v) = \sum_{jl} \left\langle (E(jl,t) - \mathbf{S}(jl,t)) \cdot \frac{1}{(Nm_j)^{1/2}} \mathbf{e}(j,\mathbf{k},v) \frac{1}{N^{1/2}} \sum_{jl} m_j^{1/2} \mathbf{e}^*(j,\mathbf{k},v) \dot{\mathbf{u}}(jl,t) \right\rangle \quad (17)$$

For Gamma point of a phonon mode, different basis atoms have the same eigenvector. Therefore,

$$\mathbf{Q}(\mathbf{k}=\mathbf{0},v) = \sum_{jl} \left\langle (E(jl,t) - \mathbf{S}(jl,t)) \cdot \frac{1}{m_j^{1/2}} \frac{1}{N} \sum_{jl} m_j^{1/2} \dot{\mathbf{u}}(jl,t) \right\rangle \quad (18)$$

Based on Eq. (18), we can know that the heat current of phonon modes contributed by Gamma point is related to the momentum of the system, $\sum_{jl} m_j^{1/2} \dot{\mathbf{u}}(jl,t)$, which is zero in the NEMD system as shown in Fig. 1, since the two ends of the system are fixed. Thus, the thermal conductivity contribution of phonon modes by the Gamma point should be zero in a steady NEMD simulation. To verify this, the heat current contribution from the Gamma point for the case of LJ Argon is shown in Fig. 7. It is clearly seen that the contribution is equal to zero, when the heat current reaches steady state.

## VII. CONCLUSIONS

In summary, we have developed a computational framework named time domain direct decomposition method, which is directly based on non-equilibrium molecular dynamics simulation and lattice dynamics theory, to characterize thermal conductivity contribution from individual phonon modes. The new TDDDM approach is validated by studying the cases of bulk LJ Argon and SW Si and comparing the results to that by SED and TDNMA from equilibrium molecular dynamics simulation. Some important parameters involved in TDDDM simulation are explicitly considered and fully discussed. The TDDDM approach inherently possesses the nature of both NEMD simulation and lattice dynamics. Comparing to EMD methods (SED and TDNMA), TDDDM has some distinct features and power in three aspects: (1) It provides direct evidence and quantitative analysis for the length dependence of thermal conductivity in NEMD simulation. We investigated the size effect of individual phonon mode contribution to the overall thermal conductivity in the TDDDM framework. We observe that the phonon modes with mean free path larger than the system size are truncated in TDDDM (NEMD) simulations, and therefore contribute little to the thermal conductivity. With system length increasing, phonons with mean free path longer than the system length are presented and then they are able to contribute to the thermal conductivity. (2) Using TDDDM, we prove that the thermal conductivity contributed from Gamma point is related to the momentum of the system, which is equal to zero for a stable system. Thereby the well-known common sense of the zero contribution of Gamma point can be rigorously proved using TDDDM. (3) We anticipate that TDDDM can be effectively used for studying phonon behaviors in nanostructures or under confined conditions, in particular when the structure size is significantly smaller than the characteristic MFP of the dominant phonons, which also cannot be realized by EMD methods. Moreover, an extension of this work, which is undergoing currently, is to apply TDDDM to deal with thermal transport across interfaces, in particular calculate phonon mode dependent transmission coefficient.

## Acknowledgements

Y.Z. thanks Mr. Tao Wang (Ruhr University Bochum) for useful discussions and providing some source codes. The fruitful discussions with Dr. Tao Ouyang (RWTH Aachen University) are also acknowledged. Simulations were performed with computing resources granted by the Jülich Aachen Research Alliance-High Performance Computing (JARA-HPC) from RWTH Aachen University under Project No. jara0127.

## Appendix A

Based on Eq. (11), TDDDM can also be used to calculate the phonon relaxation time or MFP, since we know the thermal conductivity, the group velocity and heat capacity of each phonon mode. The latter two properties can be calculated from the phonon dispersion curve by lattice dynamics. However, we have to emphasize that the number of wave vectors should be large enough, since we have to perform the discrete space Fourier transform in TDDDM. Analogously, in FDDDM the points used to do the time Fourier transform in a time interval should be adequate as well. Here, we choose the $8 \times 8 \times 400$ u.c. for LJ Argon and $4 \times 4 \times 4000$ u.c. for SW Si as examples to analyze this problem. Based on the results (FIG. A1 and FIG. A2), we can find that the trends of the relaxation time and MFP calculated using TDDDM and TDNMA are almost in the same magnitude. We try to explain the difference of relaxation time and MFP between TDNMA and TDDDM below.

Based on Eq. (1), we know that the heat current calculated using TDDDM should be equal to that using NEMD. It means that we should use large enough k mesh points in the Brillouin zone to obtain the true thermal conductivity of each phonon mode. Here, in the authors' view, the difference mainly comes from two sides. Firstly, we can only obtain the modal thermal conductivity approaching the true modal thermal conductivity since we cannot use an infinite mesh of the Brillouin zone in practice, which means the modal thermal conductivity is always larger than the true one (given that the total heat flux is same), or equivalently, the relaxation time or MFP calculated by TDDDM is larger than the true one since we use the true group velocity and heat capacity. Secondly, the relaxation time calculated using the Eq. (12) is an approximate method, and then the absolute value of the relaxation time computed using TDNMA also has some errors. However, the relative contribution of each phonon mode is more or less the same (Fig. 3 and Fig. 4), since the same phonon mode should have the same "effect" on the overall thermal transport.

**Appendix B**

As reported by L. Hu *et al.* [50], there is divergence of thermal conductivity when NEMD method is used. Furthermore, they conclude that for the isotropic materials ($C_{xx}/C_{zz} \approx 1$), the direct method leads to converged thermal conductivity for any reasonable aspect ratio (more than ~400 is needed to generate divergence). In our paper, $C_{xx}/C_{zz} \approx 1$ for both LJ Argon and SW Si and the largest aspect is 400. At the same time, our results (FIG. A3) also show the total thermal conductivity is almost independent of the cross-sectional area, when periodic boundary conditions are applied along the lateral directions, which also provides some support for the absence of divergence issue in our simulations. Therefore, we regard there is no divergence of thermal conductivity in our NEMD simulations.

**Appendix C**

Since our TDDDM scheme is established on decomposing the atomic velocity in the heat current formula onto each phonon mode, our heat current computed using TDDDM should in principal be equal to the value calculated by Eq. (1). We choose one case of LJ Argon as an example to analyze this issue. We use three different methods to calculate the heat flux flowing through the system: (1) TDDDM, i.e. the summation of all modes' contribution; (2) MD, i.e. Eq. (1) output directly from LAMMPS package; (3) time derivative of energy exchange with the external thermostats, which is the most accurate. The results are compared in FIG. A4. We find that the heat flux computed using the three methods are in good agreement with each other. There is also a noticeable difference between TDDDM and NEMD. It is possible to get such difference between TDDDM and LAMMPS itself. It is because in our TDDDM we firstly decompose the velocity into each mode to obtain the normal mode velocity and then compose all the atoms together, which implies we actually multiply the eigenvector $N^2$ times for each mode. Therefore, a small error in the eigenvector can lead to a noticeable difference in our result. For the FDDDM (in our paper serial-II), the underestimate of heat current may come from the negligible macroscopic atom diffusion (first term in Eq. (7) in our paper serial-II). Although the heat flux computed by LAMMPS itself has some variation, we think it may be due to the fact that the average times is not enough since we can find that the heat flux calculated using $J = dE/(S \cdot dt)$ is already stable.

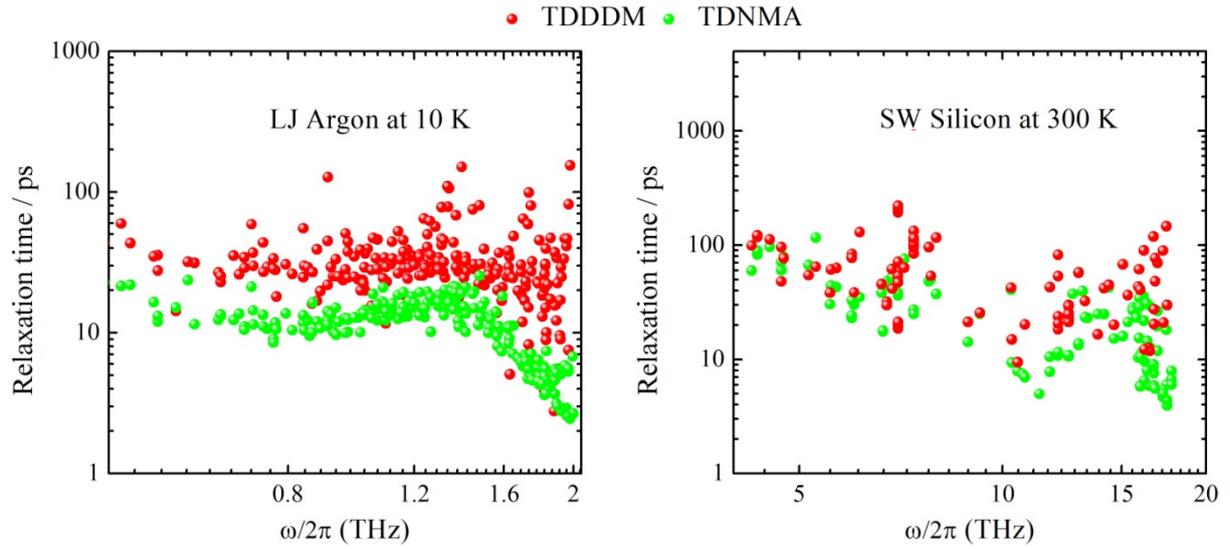

FIG. A1. Relaxation time of (left) LJ Argon and (right) SW Si calculated using TDNMA and TDDDM.

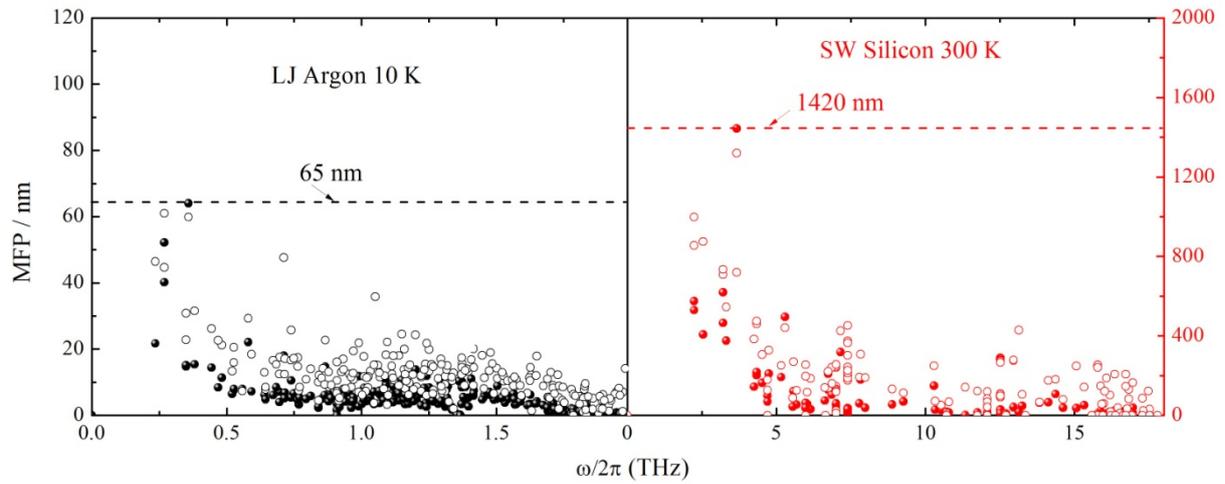

FIG. A2. Mode dependent phonon mean free path (MFP) of (left) LJ Argon and (right) SW Si calculated using TDNMA (filled circles) and TDDDM (open circles). The dashed line denotes the largest MFP in the system.

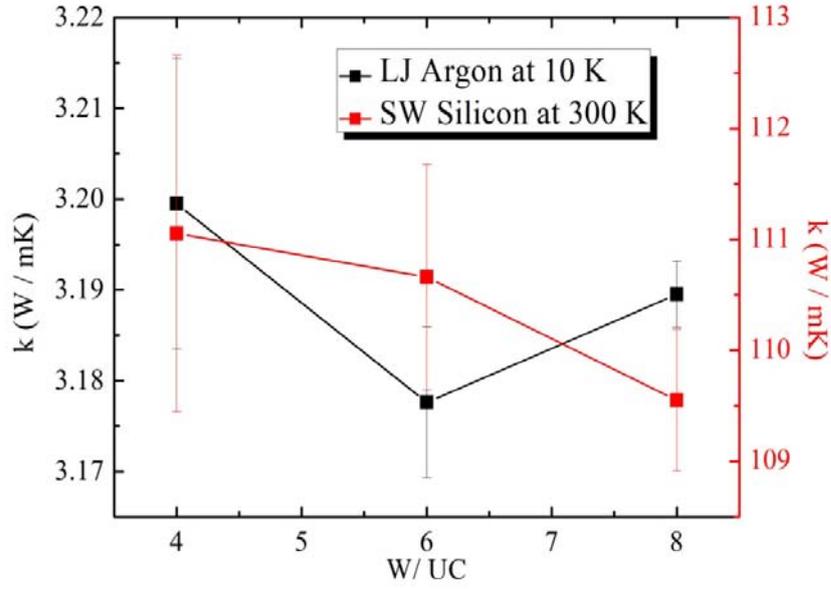

FIG. A3. Thermal conductivity of LJ Argon (left axis) and SW Si (right axis) as a function of the cross-sectional size (denoted by W, in unit of lattice constant) calculated using TDDDM method.

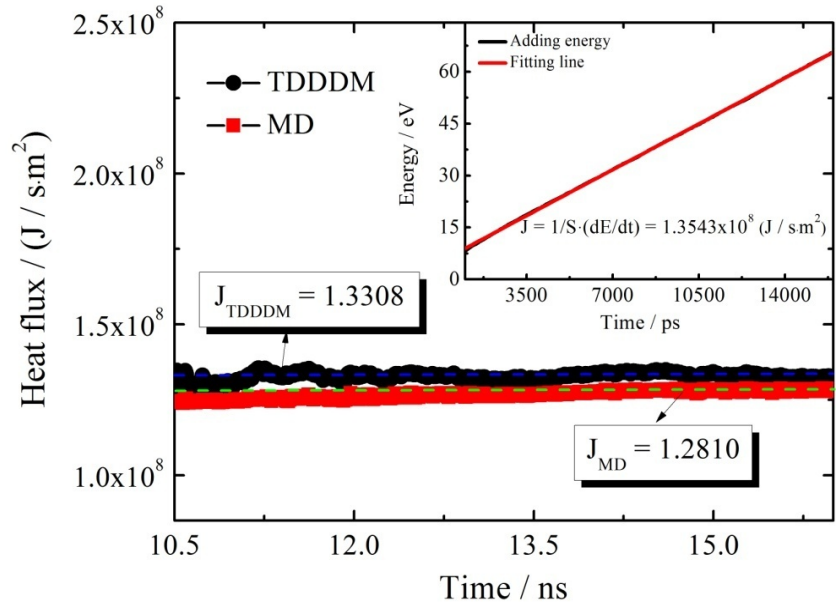

FIG. A4. The heat flux calculated by TDDDM and MD (directly output in LAMMPS using Eq. (1)). (Inset) heat flux computed from the time derivative of energy exchange with external thermostats using $J = dE / (S \cdot dt)$.

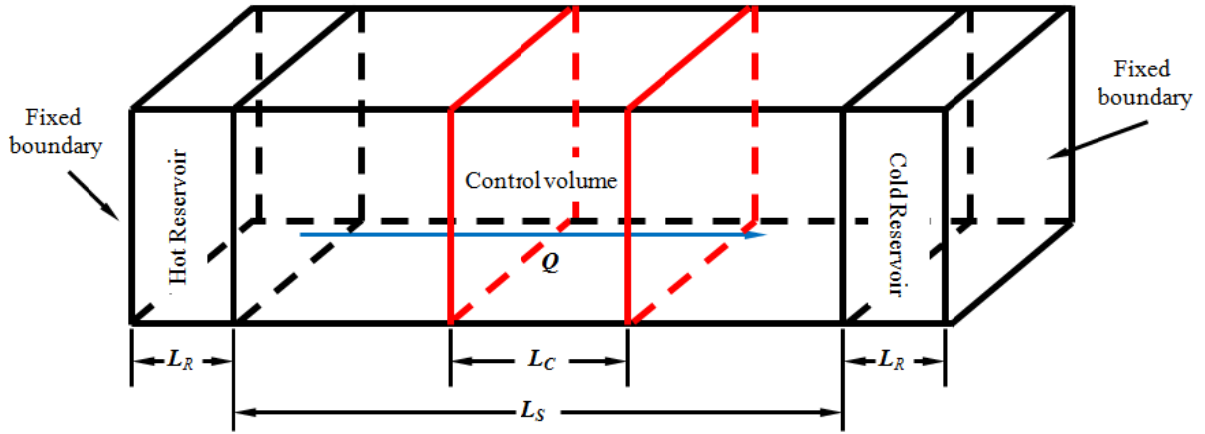

FIG. 1. (Color online) Schematic of simulation cell used in TDDDM (NEMD) simulation. $Q$ represents the heat current in the system. $L_S$, $L_C$, and $L_R$ is the total length of the system, the length of the control volume, and the length of thermostats, respectively.

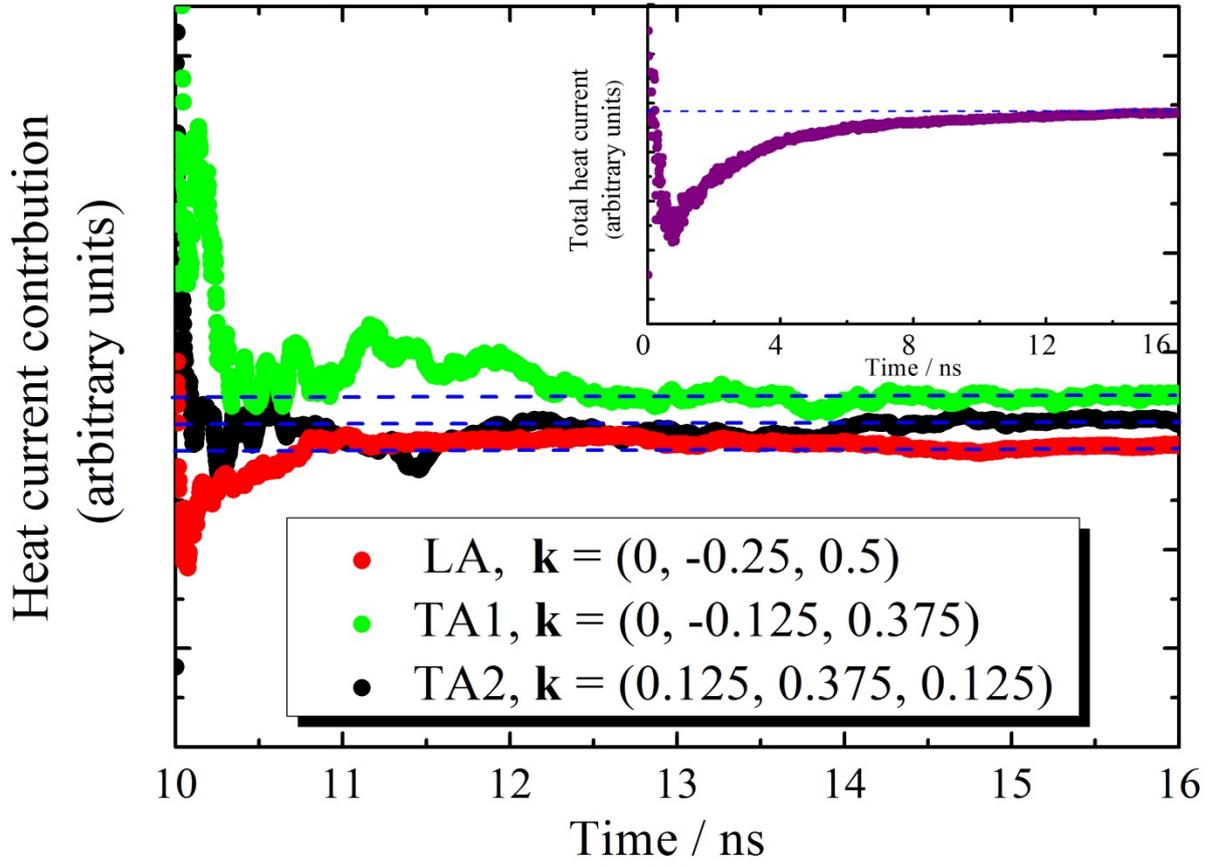

FIG. 2. (Color online) Heat current contributions of three typical individual phonon modes of LJ Argon, normalized by total predicted heat current. (Inset) Time development of total predicted heat current of the system. The dashed lines are guide for the eyes.

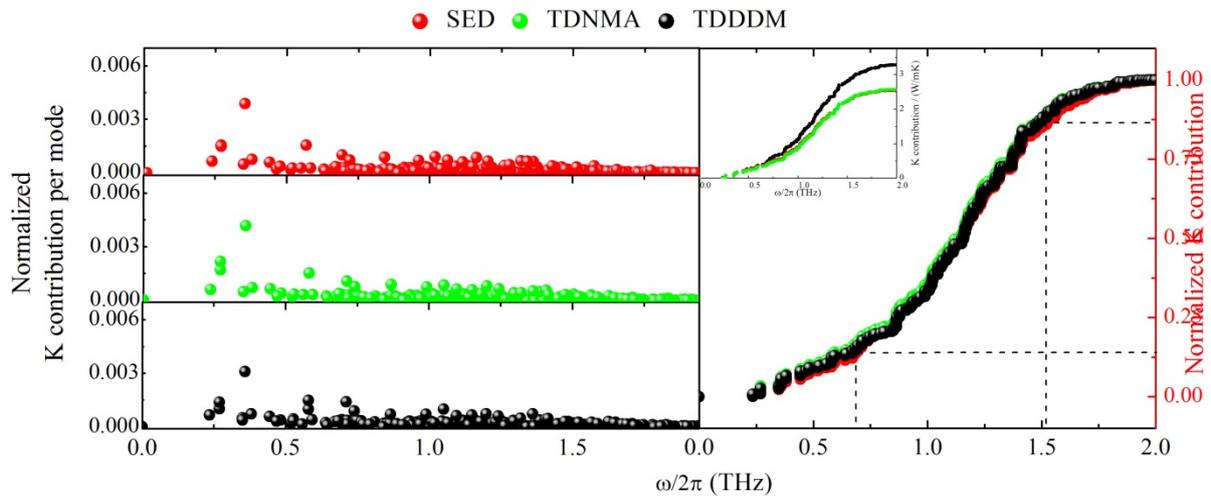

FIG. 3. (Color online) Mode dependent thermal conductivity contribution (left) and accumulative thermal conductivity as a function of mode frequency (right) for the case of LJ Argon calculated using SED, TDNMA, and TDDDM methods. The mode dependent thermal conductivity and accumulative thermal conductivity is normalized by the overall thermal conductivity predicted by each respective method. The dashed lines are guide for the eyes.

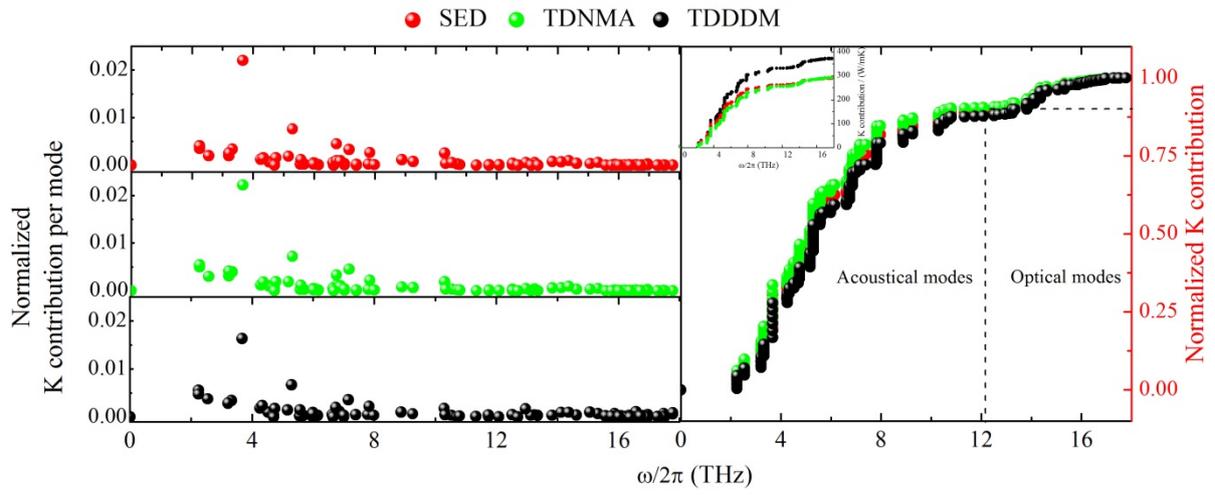

FIG. 4. (Color online) The same as Fig. 3 for the case of SW Si. The dashed lines are guide for the eyes.

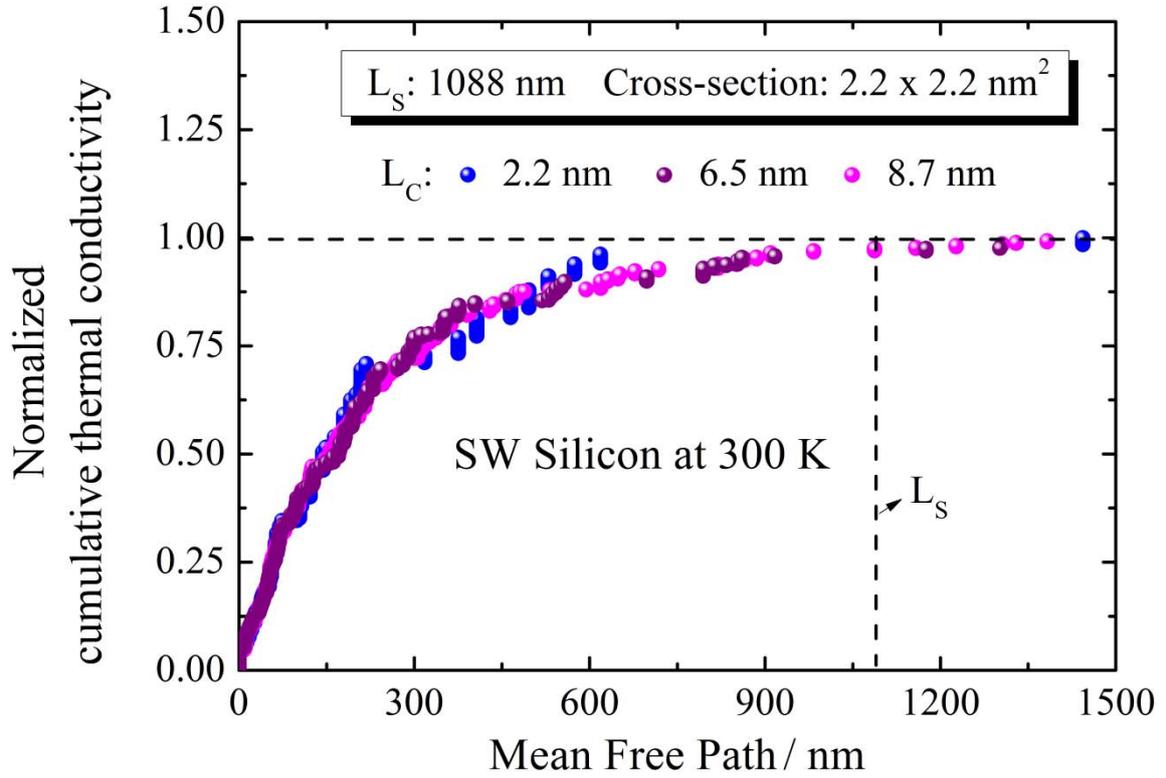

FIG. 5. (Color online) Effect of size of control volume on the mean free path (MFP) dependent accumulative thermal conductivity for SW Si in TDDDM. The accumulative thermal conductivity is computed by $k(\Lambda_0) = \sum_{\Lambda<\Lambda_0} k(\Lambda)$ and is normalized by the overall predicted thermal conductivity. $L_S$ (2000$a_{Si}$, i.e. 1088 nm) and $L_C$ represents the total length of the entire system and the length of the control volume, respectively (see definitions in Fig. 1).

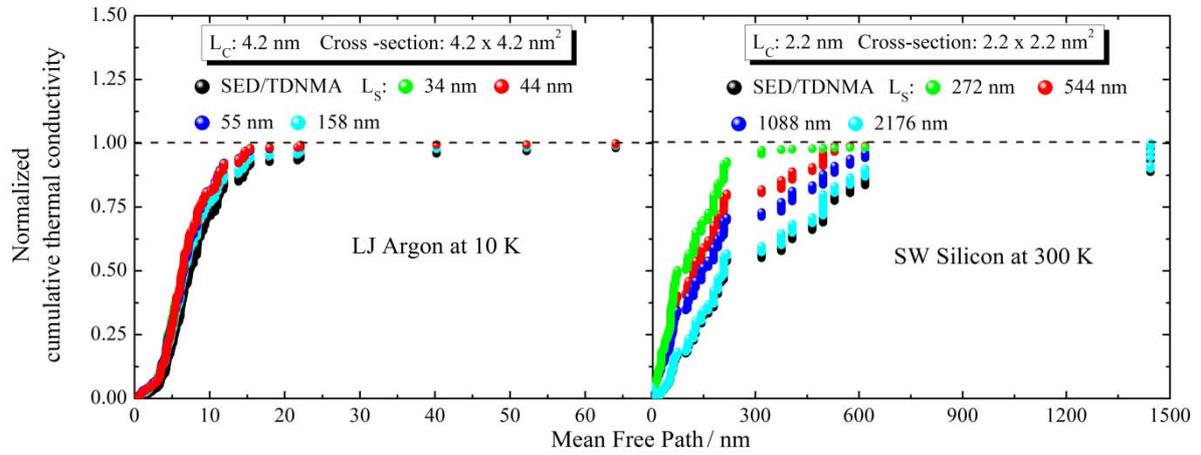

FIG. 6. (Color online) Comparison of the accumulative thermal conductivity as a function of frequency for the case of LJ Argon (left) and SW Si (right) with different total length in TDDDM. The results of SED and TDNMA are also shown for comparison with sample size of $4.2 \times 4.2 \times 4.2$ nm$^3$ (LJ Argon) and $2.2 \times 2.2 \times 2.2$ nm$^3$ (SW Si).

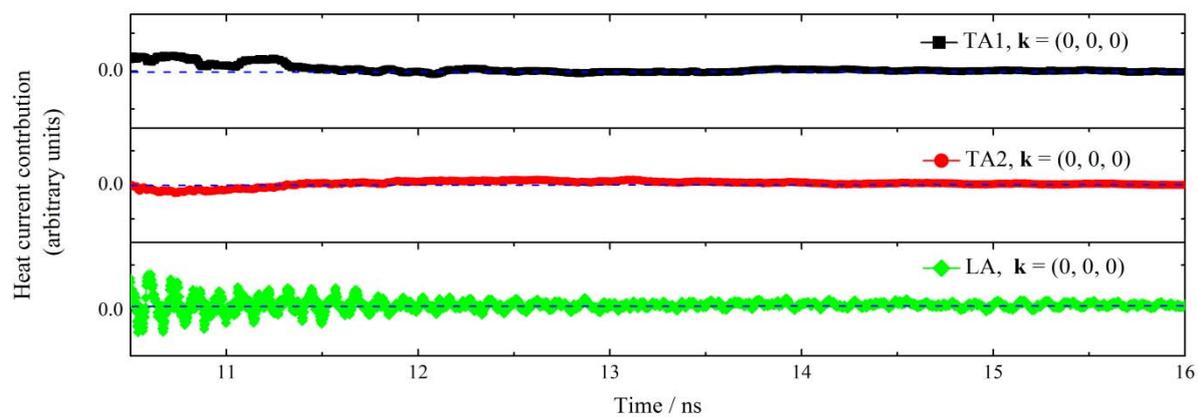

FIG. 7. (Color online) Time development of heat current contributed by the Gamma point for the case of LJ Argon predicted by the TDDDM. Dashed lines denote zero heat current and are guide for the eyes.